\documentclass[usenatbib,fleqn,twocolumn]{mnras}
\usepackage{float}
\usepackage{mathptmx}
\usepackage{amsmath,amssymb,amsfonts,amsbsy} 
\usepackage[pdftex]{graphicx} 
\usepackage[usenames,dvipsnames]{color}
\usepackage{natbib}
\usepackage{ulem}
\usepackage{threeparttable}
\DeclareSymbolFont{cmletters}{OML}{cmm}{m}{it}
\DeclareMathSymbol{v}{\mathalpha}{cmletters}{"76}

\definecolor{MyDarkBlue}{rgb}{0,0.1,0.7}
\hypersetup{pdfborder={0 0 0},colorlinks,breaklinks=true,
	urlcolor={blue},citecolor={MyDarkBlue},linkcolor={magenta}}

\title[Strong internal magnetic fields of CCOs]{On the origin of the strong internal magnetic fields of central compact objects}

\author[Ek{\c s}i \& Bak{\i}r]{Kaz{\i}m Yavuz Ek\c{s}i$^1$ and {\.I}rem Bak{\i r}$^1$ \\  
	$^1$Istanbul Technical University,
	Faculty  of Science  and  Letters,  Physics Engineering  Department,
	34469,  Istanbul, Turkey, \href{mailto:bakirir@itu.edu.tr}{bakirir@itu.edu.tr}, \href{mailto:eksi@itu.edu.tr}{eksi@itu.edu.tr}}

\begin{document}
	
	\maketitle

\begin{abstract}
Central compact objects are radio-quite young neutron stars associated with supernova remnants. They have relatively small dipole fields, $B_{\rm p} \sim 10^{10}\,{\rm G}$ as inferred from their spin parameters. X-ray observations and theoretical arguments imply the presence of stronger internal magnetic fields. We argue that the dipole fields of these objects are very close to what they had inherited from the \textit{core} of the progenitor by flux conservation and their small initial rotation frequency does not allow for the $\alpha$-process to enhance their poloidal fields. Although a full-fledged dynamo process can not proceed, relatively strong toroidal magnetic fields, $B_\phi \sim 10^{13}\,{\rm G}$, can be generated from the seed poloidal fields via the $\Omega$-effect in the proto-neutron star stage. We present a simplistic model for these processes and further speculate that the reason why these objects are born relatively slow-rotating is that they were not spun-up by acquiring angular momentum from the fallback matter.
\end{abstract}
	
	
	\section{Introduction}
	\label{sec:intro}
	
	Central compact objects (CCOs)  are young isolated neutron stars associated with supernova remnants (SNRs), but can not be classified as Crab-like rotation-powered pulsars (RPPs) or magnetars. CCOs as a class have $\sim 10$ members, the most well-known being Cas~A \citep{pav+00,cha+01} which is $\sim 350\,{\rm yrs}$ old. 
	Until recently,	CCOs were considered as radio-quiet sources with tight upper limits on their radio emission \citep{gae+00,hal+07,lu+24}.
Recent detection of radio pulsations from CCO 1E 1207.4--5209 with the MeerKAT array in the UHF band (544--1088 MHz) \citep{zha+25} demonstrates that at least some CCOs are faint radio pulsars. The X-ray emission of CCOs is thus dominated by the thermal components usually the sum of two blackbodies ($k_{\rm B} T = 0.2-0.5\,{\rm keV}$) with very small emitting radii ($\sim 0.1-4\,{\rm km}$).  
	The typical X-ray luminosity of CCOs, $L_{\rm X} \sim 10^{33}\,{\rm erg~s^{-1}}$, is  higher than their spin-down power, $L_{\rm sd} = - I \Omega \dot{\Omega}$ where $I$ is the moment of inertia of the neutron star and $\Omega$ is the angular spin-frequency \citep[see][for a review]{del17}. 
	
	The X-ray emission of a CCO is generally steady and pulsations are detected from only three of the $\sim 10$ sources to date: 1E~1207.4-5209 in G296.5+10.0 with $P\sim 424\,{\rm ms}$ \citep{zav+00}; CXOU~J185238.6+004020 in Kes 79 
	with $P \sim 105\,{\rm ms}$ \citep{got+05}; and RX~J0852.0-4622 in Puppis A with $P\sim 112\,{\rm ms}$ \citep{got09}
	where $P=2{\rm \pi}/\Omega$ is the spin period.
	
	The measurement of the period derivatives, $\dot{P}$, of the pulsating CCOs, 1E~1207.4--5209 \citep{got07,got+13a,got20}, CXOU~J185238.6+004020 \citep{hal+07,hal10a} and RX~J0852.0-4622 \citep{got+13a} imply that these objects have poloidal fields of $B_{\rm p} =(9.8,~3.1,~2.9) \times 10^{10}\,{\rm G}$, respectively. These field values are two orders of magnitude smaller than Crab-like RPPs and 4 orders of magnitude smaller than magnetars. Hence, CCOs are sometimes called ``anti-magnetars'' \citep{hal10a,got+13a}. These small magnetic dipole fields imply that the initial periods of CCOs can not be much different from their present periods. The evenly spaced strong absorption features in the X-ray spectrum of 1E~1207.4--5209 \citep{mer+02,san+02,big+03,del+04} with the fundamental at $0.7\,{\rm keV}$ indicate to a magnetic field of $8\times 10^{10}\,{\rm G}$ if interpreted as a cyclotron absorption feature \citep[see][for a more detailed account of the nature of the features]{sul+10} which is roughly consistent with the value inferred from dipole spin-down ($9.8\times 10^{10}\,{\rm G}$).
	
%\footnote{We note that, CCOs could be a heterogeneous group; for example, a once member of this group, 1E~161348--5055 at the centre of SNR RCW103 \citep{del+06}, exhibited magnetar-like activity \citep{dai+16,rea+16} and so can no longer be classified as a CCO. We, thus, exclude this object from our discussion and limit the scope of the text to those dubbed as ``anti-magnetars''.}
	
	There is some evidence that the poloidal field of CCOs is not the whole story: The high pulsed fractions ($64\%$) observed in X-rays from CXOU~J185238.6+004020 indicate that CCOs can have crustal fields much stronger than their dipole fields that leads to anisotropic heat conduction on the surface \citep{sha12,bog14}. 
	The glitch of magnitude $\Delta \Omega/\Omega \sim 10^{-9}$ observed from 1E~1207.4--5209 \citep{got18}, given that the spin-down rate of this object is much smaller than other glitching pulsars \citep{esp+11}, is also interpreted to be induced by the motion of a stronger internal field \citep{got20}. This interpretation is motivated even further by high value of $\Delta \dot{\Omega}/\dot{\Omega} \sim 0.1$ which is not compatible with pulsars, but typical with magnetars \citep{per+06,gug17,got20}. The decay of such internal fields are required also for addressing the X-ray luminosities of these objects higher than their spin-down power.

	\citet{gou+20} suggested that the peculiar emission properties of CCOs can be addressed by tangled magnetic field configurations formed in a stochastic dynamo \citep{tho01_transport} during the proto-neutron star (PNS) stage. \citet{igo+21} studied the magnetothermal evolution of this tangled crustal magnetic fields and found that all configurations can produce the complicated surface thermal patterns that consist of multiple small hot regions located at significant separations from each other. The dynamo mechanism requires that the PNS ($R \sim 40~{\rm km}$) achieves very fast rotation rates $P \sim 10~{\rm ms}$ which upon collapse to $R = 12~{\rm km}$ will lead to milisecond periods \citep{tho93}. Since the initial spin periods of CCOs are in the $\sim 100-400$~ms range, their initial periods at the PNS stage must be $1-4$~s and the dynamo mechanism with such rotation rates is not efficient \citep{nas+08}. 
	Yet another line of reasoning, the so called `field burial scenario' suggests that CCOs were born with conventional magnetic fields $B_{\rm p} \sim 10^{12}\,{\rm G}$ but suffered hypercritical accretion of fallback material left over from the supernova explosion \citep{col71,che89} which `buried' their fields \citep{ho11,vig12,ber+13,ho15,igo+16,tor+16}. This is a reincarnation of the models proposed to address the lack of a pulsar in SN87A \citep{mus95,you95,mus96,gep+99}.

	The `fossil field' hypothesis \citep{wol64,rud72} suggests that neutron stars acquire their magnetic fields from their progenitors by magnetic flux conservation. Textbook arguments which employ not the core but the whole star to collapse lead to amplification factors of $10^{10}$. 
	The radius of the iron core of a main sequence star at the brink of collapse is $\sim 3000\,{\rm km}$ \citep{suk+16}. Upon collapse of this core to a neutron star of $12\,{\rm km}$, the amplification factor will only be $(3000/12)^2 \sim 6\times 10^4$ \citep{lan21}.
	This implies that the main sequence stars even with the highest magnetic fields of $B \sim 5\times 10^4\,{\rm G}$ would form neutron stars with magnetic fields as low as $B_{\rm d}\sim 10^{10}\,{\rm G}$ upon collapse. This is an upper limit given that some fraction of the magnetic flux could be lost by turbulent magnetic reconnection.
	
	It is shown by detailed MHD simulations that although the highest magnetic fields of magnetars are generated in rapidly-rotating models, as suggested by the magnetar model \citep{dun92,tho93}, some dynamo activity is still present
	at slower rotation rates \citep{mas+20,ray+20}.
	Motivated by these, we have suggested a simple dynamo model to address both magnetar fields and that of normal pulsars \citep{bak26}. 
	In that model, for the initial periods corresponding to that of CCOs a complete dynamo action is not possible since the $\alpha$-effect is supressed. Yet, the differential rotation can still generate the relatively strong toroidal fields via the $\Omega$-effect. The purpose of the present paper is to detail these arguments to address the relatively strong internal fields of CCOs. 	
	
	The organization of the paper is as follows. In \S\ref{sec:model} we introduce the model equations for producing toroidal fields from the poloidal seed field. In \S\ref{sec:res} we show our results and in \S\ref{sec:discuss} we discuss the implications of the model for the young neutron star populations.

	\section{Model equations}
	\label{sec:model}

	\begin{table*}
		\centering
		\setlength{\tabcolsep}{3.9pt}
		\caption{SNR ages, observed periods, $P$, period derivatives, $\dot{P}$, bolometric X-ray luminosities, $L_{X,\mathrm{bol}}$, source dintances, $d$, poloidal magnetic fields, $B_{\rm p}$, toroidal magnetic fields required to power $L_{X,\mathrm{bol}}$ for $10^{3}~\rm yr$, initial fields, $B_{\rm p,0}$ and $B_{\phi,0}$, and obtained initial periods, $P_0$, and toroidal field strengths at saturation, $B_{\phi,\infty}$, of central compact objects.}
		\begin{tabular}{lccccccccccc}
			\hline
			CCO & SNR & Age & \(P\) &\(\dot{P}\) & $L_{X,\mathrm{bol}}$ & $d$ & \(B_{\rm p}\) & \(B_{\phi}^*\) &\(B_{\rm p,0}=B_{\phi,0}\)$^\dagger$ & \(P_0\)$^\dagger$ &  \(B_{\phi,\infty}\)$^\dagger$ \\
			& & (kyr) & (s) & (\(\rm s\,s^{-1}\)) & (\(10^{33}\rm erg\,s^{-1}\)) & (kpc) &($10^{10}$ G) & ($10^{13}$ G) & (G) & (s) & ($10^{13}$ G) \\
			\hline
			PSR J1210-5226$^{\rm a,~b}$  & G296.5+10.0 & \(7.0\) & \(0.424\) & \( (2.22 \pm 0.02) \times 10^{-17} \) & \(2.5\) & 2 & \( 9.8\)& \(2.15\) &\(1.2\times 10^{10}\) & \(5.0\) & \(3.9\)\\
			PSR J1852+0040$^{\rm c,~d}$ & Kes 79 & \(7.0\) & \(0.105\) & \( (8.67 \pm 0.01) \times 10^{-18} \) & \(5.3\) & 7.1 & \( 3.1\) & \(3.13\) &\(3.9\times 10^{9}\) & \(1.2\) & \(3.9\)\\
			PSR J0821-4300$^{\rm e,~f}$& Puppis A & \(3.7\) & \(0.112\) & \( (9.28 \pm 0.36) \times 10^{-18} \) & \(5.6\) & 2.2 & \( 2.9\) & \(3.22\) & \(3.6\times 10^{9}\) & \(1.3\) & \(3.8\)\\
			\hline
		\end{tabular}
		\begin{tablenotes}
			\small
			\item $Notes.$ 
			\item References. $\rm ^{a}$ \citet{zav+00,del+04}; $\rm ^{b}$ \citet{got07,got+13a,got20}; $\rm ^{c}$ \citet{got+05}; $\rm ^{d}$ \citet{hal+07,hal10a}
			$\rm ^{e}$ \citet{got09}; $\rm ^{f}$ \citet{got+13a}  
			\item $\rm ^{\dagger}$ These values are provided by this model.
			\item $\rm ^*$ These are the saturation values of the toroidal field inferred from the observed X-ray luminosities.
		\end{tablenotes}
	\label{tab:data}		
	\end{table*}
	
	The CCOs with measured period and period derivatives have inferred poloidal magnetic fields of $B_{\rm p} \sim 10^{10}~{\rm G}$. The same spin properties lead to characteristic ages $\tau_{\rm c}\equiv P/2\dot{P}$ much longer than the supernova remnant ages of these objects. This very long spin-down timescales results from the relatively small dipole fields of these objects. Thus, it can be assumed that the observed spin-periods of these objects, $P\sim 0.1-0.4~{\rm s}$, are very close to the spin periods at their birth. We thus assume in this work that the spin period at the end of the settlement of the PNS ($P_{\infty}$) is similar to the spin period at the present age.
	
	We use exactly the same model presented in our recent previous work, \citep{bak26}, which essentially follows \citet{wic+14} in many aspects. The main distinction with \citet{bak26} is that we choose the initial spin period of the PNS as $P_0 = 1.2-5~{\rm s}$ much longer than what we employed for ordinary neutron stars and magnetars. This range of initial PNS periods gives, upon settlement of the PNS to the typical neutron star radii, the observed range of the spin periods of the CCOs, $P\sim 0.1-0.4~{\rm s}$, assuming angular momentum conservation. We also adopt a lower initial shear rate as the PNS initially exhibits larger spin periods.
	
	Such long initial spin periods for the PNS do not allow for a ``full-fledged'' dynamo process to operate since slow rotation weakens the rotational constraint on convection and thus reduces the helical effect, making large scale field amplification inefficient. Thus, the $\alpha$-process that would generate the poloidal field from the toroidal one can not be effective. Yet, as we show, the $\Omega$-process that generates the toroidal field from the poloidal one via differential rotation is still effective and produces the relatively strong internal fields. \citet{bak26} suggested this `onelegged' dynamo is what generates the relatively strong internal magnetic fields of CCOs, as we detail in this work.
	
	Numerical simulations of \citet{bra09} indicate that, to be stable, the toroidal and poloidal fields should satisfy
	\begin{equation}
		a \eta^2 < \eta_{\rm p} < 0.8 \eta
		\label{eq:stability}
	\end{equation}
	where $a$ is the buoyancy factor, $\eta \equiv E/|U|$ is the ratio of the magnetic energy, $E=E_{\rm p}+E_\phi$, to the gravitational potential energy, and  $\eta_{\rm p}\equiv E_{\rm p}/|U|$ is the magnetic energy in the poloidal field scaled with the gravitational potential energy. As in \citet{bak26}, we choose the buoyancy factor as $a=10^{3}$ appropriate for neutron stars \citep{bra09}.
	
	Within the star, differential rotation twists and winds up the poloidal field lines, resulting in the generation of the toroidal field, $B_{\phi}$, which is referred as the $\Omega$-effect. The $\alpha$-effect generates the poloidal field, $B_{\rm p}$, from the toroidal field in the dynamo process. In the case of a nonrotating star, field instabilities develop over an Alfvén crossing-timescale, denoted as $\tau_{\rm A}= R/v_{\rm A}$ where $R$ is the radius of the star, $v_{\rm A} = B/\sqrt{4\pi \bar{\rho}}\simeq 10^5B_{12}~{\rm cm~s^{-1}}$ is the Alfvén velocity and $\bar{\rho}=3M/(4\pi R^3)\simeq 10^{14}~{\rm g~cm^{-3}}$ is the mean density of the star. When rotation is present, the instability growth rate decreases by a factor of $\Omega \tau_{\rm A}/(2\mathrm{\pi})$ \citep{pit85} where $\Omega$ is the stellar angular velocity. If the toroidal field does not meet the constraint given in \autoref{eq:stability}, it may diminish as a result of turbulent diffusivity with a time-scale given by 
	\begin{equation}
		\tau_{\phi}=
		\begin{cases}	
			\infty & \text { if } a \eta^2<\eta_{\rm p}~, \\
			\max \left(1, \frac{\Omega \tau_{\rm A}}{2 \mathrm{\pi}}\right) \tau_{\mathrm{A}} & \text { otherwise }~.
		\end{cases}
	\end{equation}
	When the field configuration becomes unstable, diffusion due to the turbulent diffusivity also reduces the poloidal field over a time-scale
	\begin{equation}
		\tau_{\rm p}=
		\begin{cases} 
			\infty & \text { if } \eta_{\rm p}< 0.8 \eta~, \\
			\max \left(1, \frac{\Omega \tau_{\rm A}}{2 \pi}\right)    
			\tau_{\rm A} & \text{ otherwise}~.
		\end{cases}
	\end{equation}
	The toroidal field is generated from the seed poloidal field by the $\Omega$-effect, and reduced by the turbulent diffusivity according to
	\begin{equation}
		\frac{d B_{\phi}}{dt} = q \Omega B_{\rm p} - \frac{B_\phi}{\tau_\phi}
		\label{eq:toroidal}
	\end{equation}
	where $q \equiv d\ln \Omega/d\ln r$ is the shear rate (differential rotation) \citep{bar+22}.
	
	In the dynamo process, the $\alpha$-effect amplifies the poloidal field. In addition, the poloidal field is expected to increase as the PNS shrinks, due to magnetic flux conservation. However, exact flux conservation is unlikely to hold in a turbulent environment, where the reconnection of small-scale magnetic structures can lead to an effective dissipation of magnetic energy and a reduction of the large-scale field \citep{mat+86,gold+92,eyi+13}. In neutron stars, magnetic flux transported into the crust can be further reprocessed by Hall-driven cascades toward smaller scales, which enhances subsequent Ohmic dissipation \citep{jon88,gep14,gepp14}. We parametrize the reconnection time-scale as $\tau_{\rm rec} \sim R/\epsilon v_{\rm turb}$, where $v_{\rm turb}\approx v_{\rm c}$ is the turbulent (convective) velocity \citep{mar+22}, and $\epsilon$ controls the reconnection rate. We adopt $\epsilon = 2.7 \times 10^{-4}$ following \citet{bak26}, which corresponds to a $28\%$ loss of magnetic flux for the parameter values considered in this work. Thus, defining an effective diffusive time-scale of
	\begin{equation}
		\frac{1}{\tau_{\rm p,eff}} = \frac{1}{\tau_{\rm p}} + \frac{1}{\tau_{\rm rec}}~,
	\end{equation}
	the poloidal field evolves with
	\begin{equation}
		\frac{d B_{\rm 	p}}{d t} = \alpha \frac{B_\phi}{\tau_\phi} - \frac{B_{\rm p}}{\tau_{\rm p,eff}} - 2 B_{\rm p}\frac{\dot{R}}{R}
	\label{eq:poloidal}
	\end{equation}
	\citep{bak26} where $\alpha$ is a dimensionless parameter of the so-called $\alpha$-effect, and defined by 
	\begin{equation}
		\alpha = \frac{\Delta R}{\lambda_{\rm e}} v_{\rm c} \text{min}\left(1, \text{Ro}^{-1}\right)
	\end{equation}
	where $\Delta R$ is the thickness of the convective shell, $\lambda_{\rm e}$ is the effective magnetic diffusivity, $\text{Ro}$ is the Rossby number which measures the effect of rotation \citep[see][for details]{bak26}. The final term represents the amplification of the field by magnetic flux conservation. 
	
	A PNS is born with an initial shear rate, $q_0$, which is first enhanced by collapse, but is eventually damped in time by internal stresses. This evolution can be described as a simple exponential model
	\begin{equation}
		\frac{d q}{d t} = -\frac{q}{\tau}
		\label{eq:differential}    
	\end{equation}
	where $\tau$ is the damping time-scale. The initial shear rate was assumed to be 0.4 in \citep{bak26} for magnetars where initial spin period of the PNS is two order of magnitude smaller than the values presented in this study. Here we assume a lower initial shear rate of the PNS and found that $q_0 = 0.2$ is an optimal initial shear rate for CCOs. As the PNS contracts, its moment of inertia, $I \propto R^2$ changes, amplifying the shear rate with a time-scale $\tau_{\rm I} = I/\left| \dot{I}\right|=R/2|\dot{R}|$.
	
	\begin{figure*}
		\centering
		\includegraphics[width=0.8\linewidth]{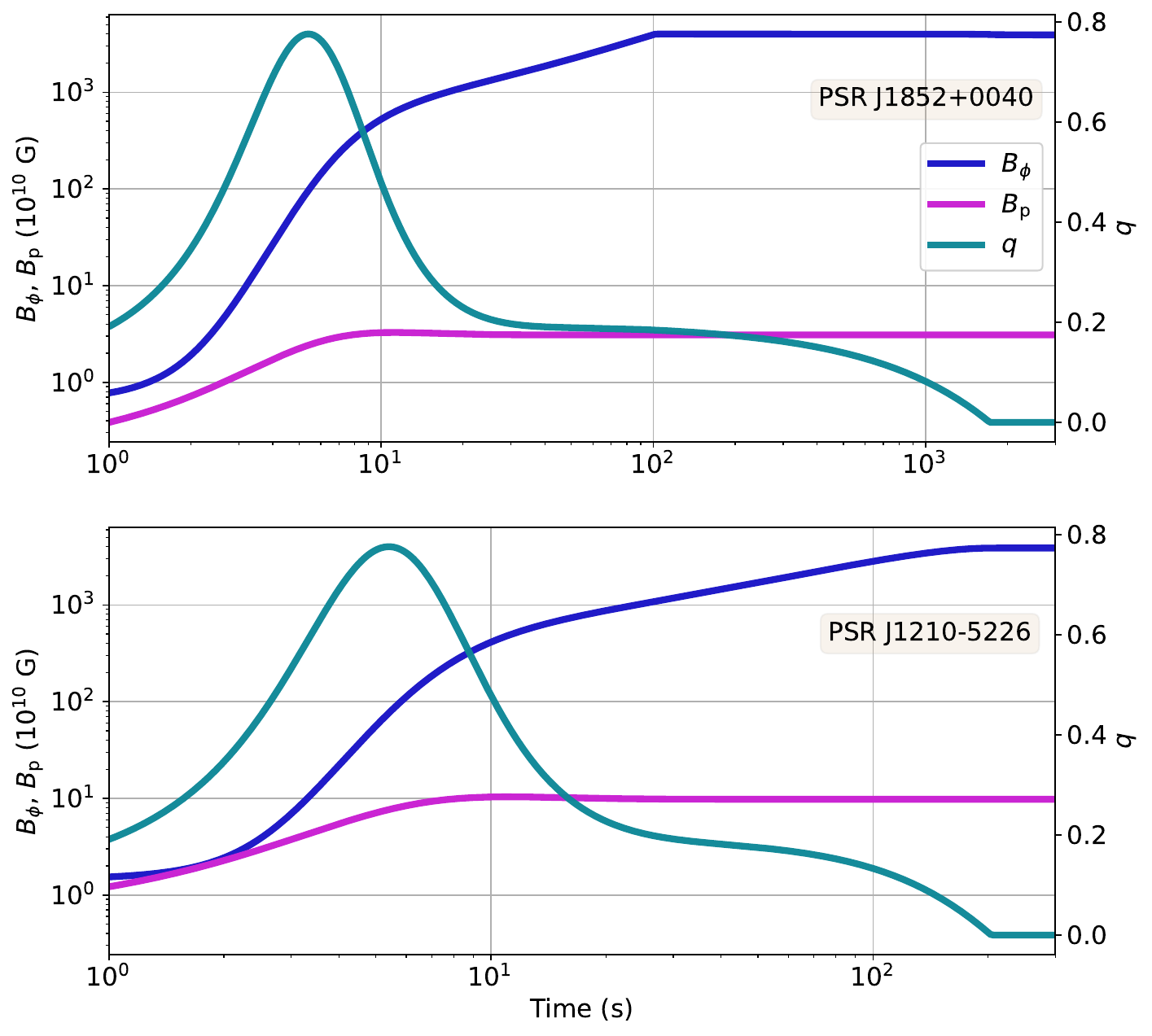}
		\caption{Evolution of the magnetic fields, $B_{\rm p}$, $B_{\phi}$, and the shear rate, $q$. (Top panel) Results for the $P_{\infty}=105\,{\rm ms}$ pulsar PSR J1852+0040 with the initial conditions of $B_{\phi}(t=0)=B_{\rm p}(t=0)= 3.9\times 10^{9}\,{\rm G}$ and $P_{0}=1.2\,{\rm s}$. (Bottom panel) Results for the $P_{\infty}=424\,{\rm ms}$ pulsar PSR J1210-5226 with the initial conditions of $B_{\phi}(t=0)=B_{\rm p}(t=0)= 1.2\times 10^{10}\,{\rm G}$ and $P_{0}=5.0\,{\rm s}$. In all models, we assumed $q_0 = 0.2$ and $a=10^{3}$.}
	\label{fig:main_fig}
	\end{figure*}
	
	The main internal stresses that eventually damp the shear rate are the viscous processes \citep{sha00,mar+22} with a viscous timescale of $\tau_\nu = R^{2}/\nu$ where $\nu$ is the internal viscosity. When a rapidly spinning PNS has strong magnetic fields, magnetorotational instability (MRI)-driven turbulence transports angular momentum effectively \citep{aki+03}, corresponding to an effective viscosity
	\begin{equation}
		\nu_{\rm MRI} = \frac{B_{\rm p} B_\phi}{4\pi\rho_{\rm crust} \left| q \right| \Omega}
	\end{equation}
	\citep{tho+05,ful+19, bar+22,mar+22} where $\rho_{\rm crust} = 10^{13}~{\rm g~cm^{-3}} \lesssim \bar{\rho}$ is the density of the PNS outermost region where MRI-driven turbulence is expected to be more relevant \citep{reb+22}.
	
 	The PNS is at first opaque to neutrinos and is anticipated to develop convective instability driven by entropy and/or leptonic gradients \citep{eps79,bur88}. We note that although the relatively long spin periods of CCOs imply weak rotational constraint on convection, turbulent convection may still provide an effective viscosity. Therefore, slow rotation mainly suppresses rotation-dependent turbulent transport mechanism of dynamo, rather than eliminating turbulent viscous dissipation. Hence, we do not neglect convective viscosity which is defined by
	\begin{equation}
		\nu_{\rm conv} \simeq \tilde{\beta} v_{\rm c} \Delta R
	\end{equation}
	\citep{mar+22} where $\tilde{\beta} = 3 \times 10^{-3}$ is the effectiveness parameter. Consequently, the total internal viscosity is given by $\nu = \nu_{\rm MRI} + \nu_{\rm conv}$. Thus, the damping time-scale of the shear rate can be written as
	\begin{equation}
		\frac{1}{\tau} = \frac{1}{\tau_\nu}-\frac{1}{\tau_I}~.
	\end{equation}
	
	During the PNS stage, the radius of the star shrinks from $R_0 \sim 40\,{\rm km}$ to $R_{\infty} \simeq 12\,{\rm km}$. The evolution of the radius of the PNS can be written as an exponential decay
	\begin{equation}
		R(t) = (R_0 - R_{\infty}) \exp\left(-\frac{t-t_0}{\tau_{ R}}\right) + R_{\infty}
		\label{eq:radius}
	\end{equation}
	where $\tau_{R}=1.3-2.3$~s depending on the EoS and the mass of the PNS \citep{bak26}. The evolution starts after $t_0 \sim 1$~s, yet we assumed $R=R_0$ during the interval $0<t<t_0$. Although this assumption is not entirely accurate, its effects are insignificant since it is a short time interval. We note that the equation given above is the solution of the differential equation $\dot{R} = -(R- R_{\infty})/\tau_{R}$
	and this form has been employed whenever it simplifies calculations.
	
	Due to the relativistic nature of the PNSs, there exists a distinction between gravitational mass and baryonic mass depending on the compactness, $GM/Rc^2$. While the baryon number and thus baryonic mass stay unchanged, a reduction in radius leads to a decrease in gravitational mass \citep{cam+17}. The evolution of gravitational mass is well-represented by an exponential function
	\begin{equation}
		M(t) = (M_0 - M_{\infty}) \exp\left(-\frac{t-t_0}{\tau_{M}}\right) + M_{\infty}
	\end{equation}
	with $3.4~{\rm s}<\tau_{M}<4.5~{\rm s}$ \citep{bak26}. Therefore, we analyzed the PNS mass evolution as governed by $\dot{M}=-(M-M_{\infty})/\tau_{M}$, which results in approximately a $10\%$ variation in mass. We again assumed that the mass is constant for $0<t<t_0$.
	
	As a result of the conservation of angular momentum, the angular velocity of the contracting PNS requires an increment as
	\begin{equation}
		\frac{d\Omega}{dt} = - \frac{\dot{I}}{I} \Omega~.
		\label{eq:omega}
	\end{equation}
	External torques ---due to electromagnetic emission, gravitational waves, and neutrino-driven winds--- would tend to spin the star down, yet they are negligible especially in the case of CCOs. Thus, the change in the moment of inertia is the dominant factor and increases the angular velocity of the star.

	\section{Results}
	\label{sec:res}
	%******************************************************************************************
	%******************************************************************************************

	We have solved the set of differential equations of the previous section with the initial conditions suitable for producing the spin periods and magnetic fields of CCOs. We adopted $M_0 = 1.55 M_\odot$ and $R_0 = 40\,{\rm km}$ as typical values for a PNS, and the final configuration is characterized by $M_\infty = 1.46 M_\odot$ and $R_\infty = 12\,{\rm km}$. We assumed $q_0=0.2$ and $a=10^{3}$. The initial angular velocities are obtained by applying angular momentum conservation to each CCO, ensuring that, once they settle, they exhibit the specific rotational periods listed in \autoref{tab:data}. The results are shown in \autoref{fig:main_fig}.

	Our results infer that PSR J1852+0040, which has $P_{\infty}=105\,{\rm ms}$, begins its evolution rotating with an initial period of $P_0 = 1.2\,{\rm s}$ at the PNS stage ($R=40$~km). For the seed fields of $B_{\phi}(t=0)=B_{\rm p}(t=0)= 3.9\times 10^{9}\,{\rm G}$, its poloidal field reaches its saturation value close the inferred dipolar field at late times from spin properties. The saturation value of the toroidal field is $B_{\phi,\infty}=3.9\times 10^{13}\,\rm G$. 
	
	For PSR J0821-4300 with a final spin period of $P_{\infty}=112\,{\rm ms}$, the model favors a very similar early-time setup as expected. It initially rotates with $P_{0}=1.3\,{\rm s}$. In this case, slightly weaker seed fields of $B_{\phi}(t=0)=B_{\rm p}(t=0)= 3.6\times 10^{9}\,{\rm G}$ leads to the present day magnetic field. It has $B_{\phi,\infty}=3.8\times 10^{13}\,\rm G$ at saturation. Because the spin parameters of this source is very similar to PSR J1852+0040, we do not show the field, differential rotation and spin evolution of this source in the figure, but just list its parameters in \autoref{tab:data}.
	
PSR J1210-5226 is the slowest rotator among the three CCOs we study here. We found that it requires a comparatively longer initial spin period, $P_{0}=5.0\,{\rm s}$. We find $B_{\phi}(t=0)=B_{\rm p}(t=0)= 1.2\times 10^{10}\,{\rm G}$ yields the appropriate saturation behavior, with $B_{\phi,\infty}=3.9\times 10^{13}\,\rm G$. 

\begin{figure}
	\centering
	\includegraphics[width=0.95\linewidth]{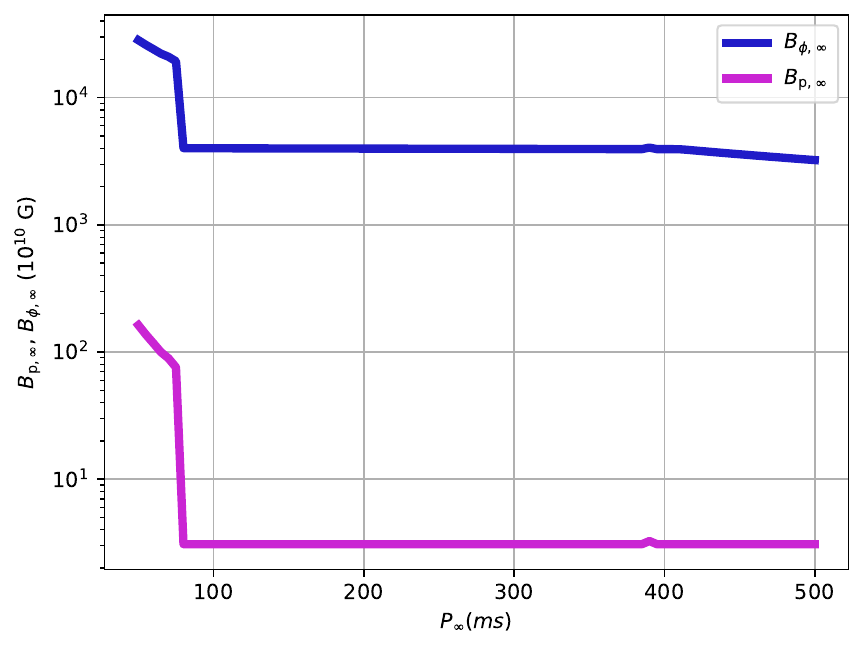}
	\caption{Effect of the final rotational period, $P_\infty$, on the saturation values of poloidal field, $B_{\rm p,\infty}$ and toroidal field $B_{\phi,\infty}$. Here, the initial fields are $B_\phi (t=0)=3.9\times10^{9}\,\rm G$ and $B_{\rm p} (t=0)=3.9\times10^{9}\,\rm G$.}
	\label{fig:main_fig3}
\end{figure}
	
We emphasize that the enhancements in the poloidal fields seen in \autoref{fig:main_fig} are \textit{not} due to the $\alpha$ effect but solely due to flux conservation during the collapse of the PNS to the typical neutron star radius.  However, the toroidal field components are amplified to strengths of $ \gtrsim 10^{13}\,\rm G$ by $\Omega$-action.
	   
\begin{figure*}
		\centering
		\includegraphics[width=0.95\linewidth]{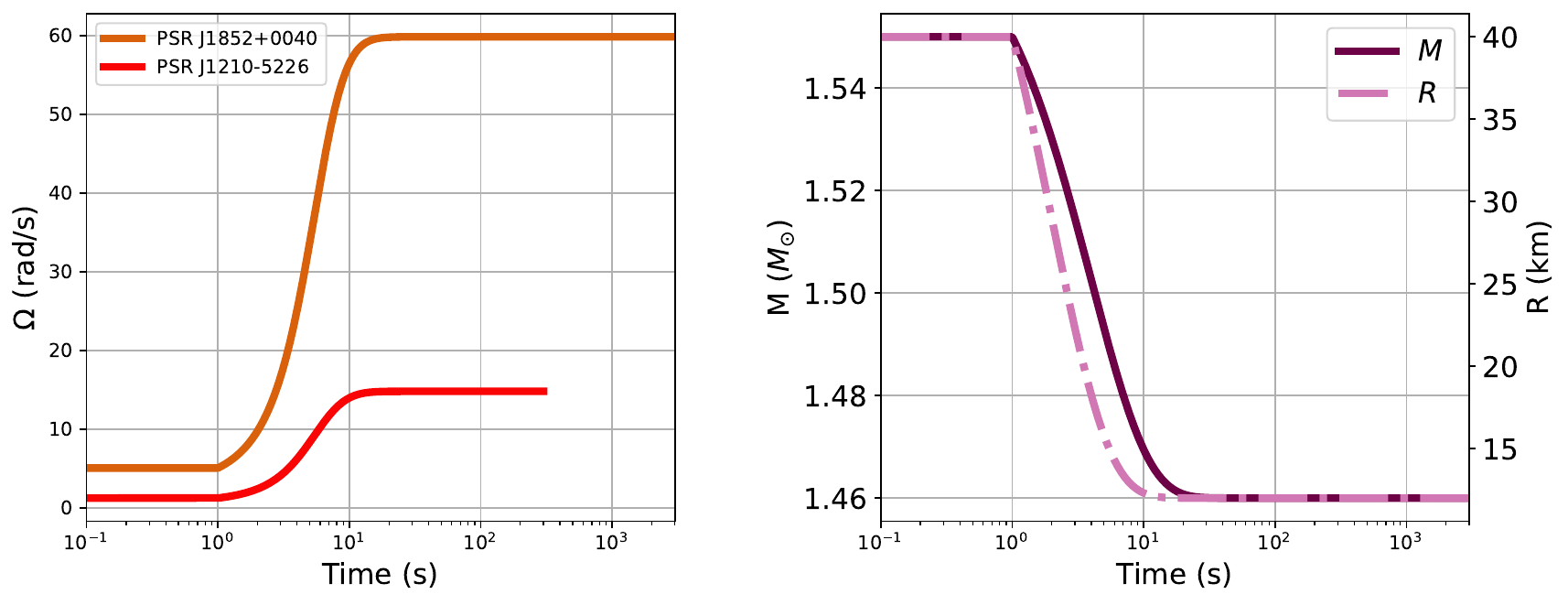}
		\caption{ (Left panel) Evolution of the angular velocity, $\Omega$, for PSR J1852+0040  and PSR J1210-5226. (Right panel) Evolution of mass, $M$, and radius, $R$, for three CCOs.}
		\label{fig:main_fig2}
	\end{figure*}

We also analyzed the effect of the final rotational period $P_\infty$ on the results by varying it within the interval of $50-500\,\rm ms$. Results depicted in \autoref{fig:main_fig3} demonstrate that when $P_\infty \leq 84\,\rm ms$, field strengths at saturation are weaker for longer rotational periods. However, for $P_\infty \geq 85\,\rm ms$, the saturation value of the poloidal field becomes  independent of $P_\infty$. This behavior indicates that in the faster rotation regime $\left(P_\infty \leq 84\,\rm ms\right)$, $\alpha$-effect is operative in the dynamo process and can enhance poloidal field above the value attainable through flux conservation alone. By contrast, beyond $P_\infty = 85\,\rm ms$, it becomes ineffective and poloidal field is no longer sustained by dynamo action, so the final poloidal field is determined predominantly by flux conservation during the stellar contraction. Similarly, the saturation value of the toroidal field is also constant over some range, but beyond $P_\infty = 407\,\rm ms$, it shows a slight decline. Nevertheless, the results indicate that the $\Omega$-effect remains operative across the entire range of rotational periods considered and generate toroidal field. Time evolution of the angular velocity, $\Omega$, mass, $M$, and radius, $R$, for PSR J1852+0040  and PSR J1210-5226 in \autoref{fig:main_fig2}.

	%******************************************************************************************
	%*****************************************************************************************
	\section{Discussion}
	\label{sec:discuss}
	%******************************************************************************************
	%******************************************************************************************

	We have presented a simple ``onelegged'' dynamo model for CCOs in which the toroidal fields are enhanced by the $\Omega$-effect while the $\alpha$-effect is suppressed owing to the the slow initial periods of these objects ($P \sim 0.1-0.4~{\rm s}$. The poloidal fields of CCOs are thus consistent with the fossil field hypothesis.
	 The generation of the toroidal fields from the poloidal ones by differential rotation by the $\Omega$-effect resulted with $B_\phi \simeq (3-4) \times 10^{13}~{\rm G}$ which is three orders of magnitude stronger than their poloidal fields inferred from their spin parameters. This all is consistent with the observations which suggest that CCOs, dubbed as ``anti-magnetars' \citep{hal10a,got+13a}, have much stronger internal fields than their inferred dipole fields \citep{sha12,bog14,got18,got20}. Moreover, these strong internal fields can be converted to dipole and multipole field components later in the lifetime of the CCOs as in \citet{igo+21}.
	
	The question then naturally arises why these objects start their lives with a spin relatively slower than the normal pulsars and magnetars. According to a recent view, the neutron stars get their spin by interacting with the supernova fallback matter, soon after their birth \citep{jan+22}. The PNS reaches the required spins for magnetar formation by interaction with the fallback matter \citep{bar+22}. This line of reasoning would imply that CCOs are those neutron stars which had the least interaction with the fallback matter or that the effect of one interaction was soon canceled by another. Since they were not accelerated to short spin periods at birth, the convection and the resulting $\alpha$-effect was supressed, and their poloidal fields were not enhanced above what they could get from the core of the progenitor via flux conservation. 
	
	Our findings demonstrate that the slow initial rotation rates significantly influence the magnetic field configurations of these neutron stars. Due to these relatively long initial spin periods, the $\alpha$-mechanism, previously shown to be effective in amplifying the poloidal magnetic fields of rotationally powered pulsars \citet{bak26}, does not operate in CCOs. Nevertheless, differential rotation in PNSs can still effectively activate the $\Omega$-mechanism, generating the toroidal component from the inherited fossil poloidal magnetic.
	
We determined the initial parameters required to obtain the surface field strengths of specific CCOs given in \autoref{tab:data}. We confirm that the initial magnetic fields of CCOs are inherited from their progenitor through magnetic flux conservation. However, it is evident by the results depicted \autoref{fig:main_fig} that the differential rotation can strengthen the fossil toroidal field in CCOs significantly up to $B_{\phi} \gtrsim 10^{13}\,{\rm G}$, while poloidal field remains relatively weak throughout the evolution with saturation values of $B_{\rm p}\sim 10^{10}\,{\rm G}$. The toroidal field strengths obtained by our model agree with the order-of-magnitude estimate of toroidal fields in canonical radio pulsars inferred by \citet{gug17}.

The magnetic energy stored in such fields, $E_{\rm mag} \sim B_\phi^2 R^2 \Delta R_\phi/2 \simeq 3 \times 10^{44}$~erg for $B_\phi \sim 4\times 10^{13}$~G, where $\Delta R_\phi= 0.2R$ is the size of the toroidal field region \citep{lan12}. This is sufficient to power the observed X-ray luminosities $L_{\rm X} \sim 10^{33}$~erg~s$^{-1}$ only for $\sim 10^3$ years. Thus the thermal emission of CCOs with ages of $\gtrsim 10^3$ yrs should arise from the cooling of the initial thermal content rather than field decay.

Below are further implications of this study. 

\subsection{Implications for the putative neutron star in SN87A}

The absence of a detected pulsar in SN 1987A has been a long-standing puzzle, with various explanations proposed including the field burial scenario \citep{mus95, mus96, gep+99} and the possibility that a black hole formed instead \citep{gra+05}. Our model offers an alternative interpretation: if a neutron star formed in SN 1987A and experienced minimal or no spin-up from fallback matter, it would have been born with a weak dipole field $B_{\rm p} \sim 10^ {10}\,{\rm G}$, similar to the known CCOs. Such a weak field would render the pulsar extremely faint in radio, consistent with the non-detection despite deep searches \citep{gra+05,zan+14}. The magnetospheric high energy luminosity of the source would also be in the low-B branch of \citet{oge04}. Unlike the field burial scenario which requires hypercritical accretion to suppress the initially high surface field, our model suggests the neutron star simply inherited a weak poloidal field from its progenitor core via flux conservation and never underwent significant dynamo amplification due to slow initial rotation ($P_0 \sim 1-5$~s). The differential rotation in the proto-neutron star phase would nevertheless generate a toroidal component of $B_\phi \sim 10^{13}$~G. Nearly 40 years post-explosion, any differential rotation would have been completely damped and the magnetic field configuration settled. The decay of this internal toroidal field could contribute to residual thermal emission, potentially contributing to the compact source observed in recent JWST observations \citep{bou+24}, though disentangling this from other emission components remains challenging. The predicted weak dipole field also implies minimal rotational energy loss, meaning the neutron star should remain relatively warm for extended periods, with cooling dominated by neutrino emission from the interior rather than being accelerated by magnetic field decay at the surface.

\subsection{Implication for the CCO in Cassiopeia A}

Cassiopeia A presents an interesting test case for our model, as it is the youngest known CCO at approximately 350 years old and has been extensively studied across multiple wavelengths. Unlike the three CCOs discussed in \S\ref{sec:res}, no pulsations have been detected from the Cas A despite extensive searches \citep{mur02}, leaving its spin period and magnetic field strength unconstrained by timing observations. Our model would predict that if Cas A was born through the same mechanism as other CCOs---minimal spin-up from fallback matter---it should have been born with $P_0 \sim 1-5$~s at the proto-neutron star phase and developed an internal toroidal field of $B_\phi \sim 10^{13}$~G through the $\Omega$-process while its poloidal field remained at $B_{\rm p} \sim 10^{10}\,{\rm G}$, the flux-conserved value inherited from the progenitor. Such a weak dipole field could explain the non-detection of pulsations: the corresponding spin-down luminosity would be extremely low, potentially rendering any pulsed emission undetectable with current instrumentation. The relatively young age of Cas A makes it particularly valuable for testing magnetic field decay models: at only 350 years, the contribution of the toroidal field decay to the thermal luminosity should be decreasing. The observed long-term decline in the thermal X-ray luminosity of the Cas A CCO, reported by \citet{hei10} has been interpreted as evidence for cooling of the neutron star core \citep[but see][]{pos+13}, potentially indicating enhanced neutrino cooling from exotic processes such as neutron superfluidity or quark matter. However, an alternative or complementary explanation within our framework is that we are witnessing the final phase of internal toroidal field decay, with the field reconfiguration depositing heat in the crust at a declining rate. The timescale of the observed cooling ($\tau_{\rm cool} \sim 100-300$~yrs; \citet{ho09}) could constrain the magnetic diffusivity in the crust and the initial toroidal field strength. Deep monitoring of the Cas A CCO's thermal emission over the coming decades, combined with continued searches for pulsations, could test whether the internal field strength predicted by our model is consistent with the observed thermal evolution and potentially reveal the spin properties of this youngest known CCO.

\subsection{Population-level implications}

Our model implies a trimodal distribution of neutron star magnetic field strengths determined primarily by the degree of interaction with fallback matter shortly after birth. CCOs represent the ``no spin-up'' channel where negligible angular momentum transfer from fallback matter results in slow initial rotation ($P_0 \sim 1-5$~s at the proto-neutron star phase), insufficient to drive the $\alpha$-mechanism. These objects retain weak dipole fields $B_{\rm p} \sim 10^{10}\,{\rm G}$ close to the values obtained by flux-conservation but develop moderate internal toroidal fields $B_\phi \sim 10^{13}$~G through differential rotation. Normal rotation-powered pulsars occupy an intermediate regime where moderate fallback spin-up yields $P_0 \sim 50-300$~ms, enabling partial $\alpha$-effect operation and producing $B_{\rm p} \sim 10^{12}-10^{13}\,{\rm G}$, as detailed in \citep{bak26}. Magnetars result from strong fallback spin-up to $P_0 \sim 1-10$~ms, allowing full dynamo action with both $\Omega$ and $\alpha$ processes operating efficiently, generating fields $B_{\rm p} \sim 10^ {14}-10^ {15}\,{\rm G}$ \citep{bar+22}. This framework naturally explains the apparent gap in the observed distribution between CCO fields ($B_{\rm p} \sim 10^{10}\,{\rm G}$) and typical pulsar fields ($B_{\rm p} \sim 10^{12}\,{\rm G}$) as a reflection of the bimodal nature of fallback interactions: either the neutron star couples minimally to fallback matter (producing CCOs) or it couples efficiently and spins up substantially (producing normal pulsars). The relative birth rates of these populations should correlate with supernova explosion properties, particularly the amount and angular momentum distribution of fallback matter, which in turn depend on progenitor structure and explosion energy \citep{che89, col71}. The fact that only $\sim$10 CCOs are known compared to $\sim$4000 normal pulsars does not immediately suggest that the ``no spin-up'' scenario is relatively rare, but that those objects are systematically under-detected due to their weak radio emission.

\subsection{Future observational tests}

Our model makes several testable predictions that can be verified with future observations. First, the predicted internal toroidal fields $B_\phi \sim 10^{13}$~G should manifest in the glitch behavior of CCOs: the magnitude and recovery timescales of glitches should differ from those of normal pulsars due to the different internal magnetic field geometry. The glitch observed in 1E 1207.4-5209 with $\Delta \Omega/\Omega \sim 10^{-9}$ and $\Delta \dot{\Omega}/\dot{\Omega} \sim 0.1$ \citep{got18,got20} is consistent with magnetic field line motion in a strong internal field; future glitches from this or other CCOs could be analyzed to extract constraints on $B_\phi $. Second, long-term thermal monitoring of Cas A should reveal cooling curves consistent with the decay of $\sim 10^{13}$~G toroidal fields.  Third, the recent radio detection of 1E 1207.4-5209 with MeerKAT \citep{zha+25} suggests that deeper radio surveys, particularly at UHF frequencies where beaming geometry and plasma effects may be more favorable, could detect additional CCOs. The radio luminosity and spectral properties of CCOs should differ systematically from normal pulsars due to their much weaker $B_{\rm p} \sim 10^{10}$~G surface fields. Finally, high-pulsed-fraction observations in X-rays, as seen in CXOU J185238.6+004020 \citep{sha12}, should become more common as X-ray timing capabilities improve; the spatial pattern of hot spots on CCO surfaces, driven by anisotropic heat conduction in strong crustal magnetic fields, can constrain the geometry and strength of $B_\phi$. 
	
\section*{Acknowledgements}
	
KYE acknowledges support from the Scientific and Technological Research Council of Turkey (TÜBİTAK) through project number 118F028. We gratefully thank Erbil Gügercinoğlu for his valuable comments and fruitful discussions.

\section*{Data availability}
No new data were analysed in support of this paper.
	
%\nocite{*}
\footnotesize{
\bibliographystyle{mnras}
\bibliography{cco.bib} 
}
\end{document}